\documentclass[%
 reprint,
superscriptaddress,
 amsmath,amssymb,
 aps,
prx,
]{revtex4-2}

\usepackage{comment}
\usepackage{graphicx}
\usepackage{dcolumn}
\usepackage{bm}
\usepackage[version=4]{mhchem}
\usepackage{xcolor}
\usepackage{hyperref}

\begin{document}

\preprint{PG and SG in RIXS}

\title{Charge response function probed by resonant inelastic x-ray scattering: \\ the signature of electronic gaps of YBa$_2$Cu$_3$O$_{7-\delta}$ }

\author{Giacomo\, Merzoni}
 \email[e-mail ]{giacomo.merzoni@polimi.it}
 \affiliation{Dipartimento di Fisica, Politecnico di Milano, piazza Leonardo da Vinci 32, I-20133 Milano, Italy}
 \affiliation{European XFEL, Holzkoppel 4, Schenefeld, D-22869, Germany}

\author{Leonardo\, Martinelli}%
\altaffiliation[current address ]{Physik-Institut, Universität Zürich, Winterthurerstrasse 190, CH-8057 Zürich, Switzerland}
\affiliation{Dipartimento di Fisica, Politecnico di Milano, piazza Leonardo da Vinci 32, I-20133 Milano, Italy}

\author{Lucio\, Braicovich}
\affiliation{Dipartimento di Fisica, Politecnico di Milano, piazza Leonardo da Vinci 32, I-20133 Milano, Italy}
\affiliation{ESRF, The European Synchrotron, 71 Avenue des Martyrs, CS 40220, F-38043 Grenoble, France}

\author{Nicholas B.\, Brookes}
\affiliation{ESRF, The European Synchrotron, 71 Avenue des Martyrs, CS 40220, F-38043 Grenoble, France}

\author{Floriana\, Lombardi}
\affiliation{
Quantum Device Physics Laboratory, Department of Microtechnology and Nanoscience, Chalmers University of Technology, SE-41296 Göteborg, Sweden
}

\author{Francesco\, Rosa}
 \affiliation{Dipartimento di Fisica, Politecnico di Milano, piazza Leonardo da Vinci 32, I-20133 Milano, Italy}

\author{Riccardo\, Arpaia}
\affiliation{
Quantum Device Physics Laboratory, Department of Microtechnology and Nanoscience, Chalmers University of Technology, SE-41296 Göteborg, Sweden
}

\author{Marco\, Moretti Sala}%
\affiliation{Dipartimento di Fisica, Politecnico di Milano, piazza Leonardo da Vinci 32, I-20133 Milano, Italy}

\author{Giacomo\, Ghiringhelli}%
\email[e-mail ]{giacomo.ghiringhelli@polimi.it}
\affiliation{Dipartimento di Fisica, Politecnico di Milano, piazza Leonardo da Vinci 32, I-20133 Milano, Italy}
\affiliation{CNR-SPIN, Dipartimento di Fisica, Politecnico di Milano, I-20133 Milano, Italy}

\date{\today}

\begin{abstract}
In strongly correlated systems the complete determination of the dynamical susceptibility $\chi(\textbf{q}, \omega)$ is of special relevance because of the entwinement of the spin and charge components. Although Resonant Inelastic X-Ray Scattering (RIXS) spectra are directly related to both the charge ($\chi''_c(\textbf{q}, \omega)$) and the spin ($\chi''_s(\textbf{q}, \omega)$) contributions, only the latter has been extensively studied with RIXS so far.  Here we show how to extract from RIXS spectra of high-$T_c$ superconducting cuprates relevant properties of $\chi''_c$, such as the presence of the superconducting gap and of the pseudogap. In particular, we exploit the temperature dependence of the Cu L$_3$ edge RIXS spectra of underdoped YBa$_2$Cu$_3$O$_{7-\delta}$ at specific wave-vectors \textbf{q}. The signature of the two gaps is given by the departure of the low energy excitation continuum from the Bosonic thermal evolution. This approach can be immediately used to investigate systematically the nature of the pseudogap in cuprates, thereby taking advantage of the RIXS technique that does not suffer the limitations of surface-sensitive electron spectroscopies. Its extension to other interesting materials is foreseen. 

\begin{description}
\item[DOI]

\end{description}
\end{abstract}

\maketitle


\section{\label{sec:level1}Introduction}

The dynamical susceptibility $\chi(\textbf{q}, \omega)$ encodes the whole response of a system to an electromagnetic perturbation and its experimental determination would entail a complete knowledge of the dynamical properties of a material. In quantum materials the complete determination of the dynamical susceptibility $\chi(\textbf{q}, \omega)$ is of special relevance because of the entwinement of the spin and charge degrees of freedom. Scattering techniques are effective at its experimental determination as they directly measure the dynamical structure factor $S_T(\textbf{q},\omega)$ \cite{{Schulke},{Vig}}, which is related to $\chi''(\textbf{q},\omega)$ through the fluctuation-dissipation theorem \cite{Schulke}. Nonetheless, all these techniques present shortcomings: optical spectroscopies are restrained to very small \textbf{q} values and have an indirect access to the spin component; electron energy loss spectroscopy (EELS) is hampered by practical issues, such as surface sensitivity; inelastic x-ray scattering (IXS) and inelastic neutron scattering (INS) are mostly sensitive to one component (charge and spin, respectively) and suffer from particularly small cross-sections. Moreover, due to the poor contribution of valence electrons in $S_T(\textbf{q},\omega)$, in the low energy region the spectra are dominated by sharp phonon peaks and the signal from particle-hole ($e$-$h$) excitations falls often below the detection threshold.

More recently resonant inelastic x-ray scattering (RIXS) has emerged for its capability to access multiple degrees of freedom. Indeed, RIXS was used to probe the orbital and magnetic character of cuprates\cite{{Moretti},{GhiringhelliDD},{Schlappa},{BraicovichMag},{LeTacon},{PengInfluence}}, to unveil the presence of charge order and charge density fluctuations in all cuprates families \cite{{GhiringhelliCDW},{ArpaiaUNO}}, and to determine the momentum dependent electron-phonon coupling \cite{{Rossi},{BraicovichPhon},{PengMartinelli}} and its interplay with charge order\cite{{chaix2017dispersive},{lee2021spectroscopic}}. By RIXS it is possible to widely explore the reciprocal space with high momentum resolution and good sensitivity to the bulk of the material. The rapid improvement of energy resolution \cite{{Brookes},{Zhou},{Dvorak}} and the good understanding of cross sections \cite{Ament} allows the assignment of spectral features with high degree of confidence.

\begin{figure}
\centering
\includegraphics[scale=0.39]{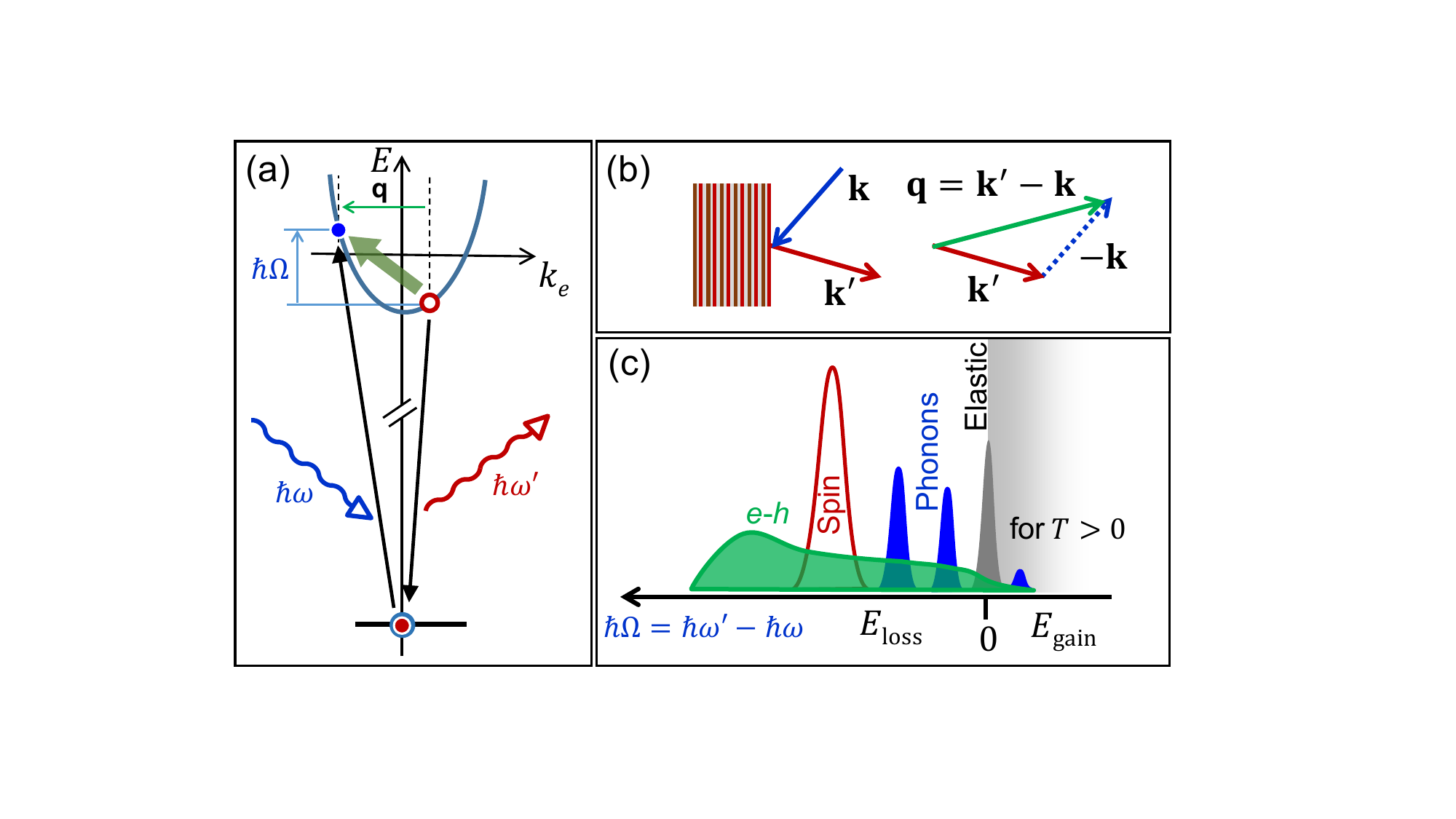}
\caption{\label{cartoon RIXS} The RIXS process and the charge susceptibility. (a) Sketch of the particle-hole excitations as they are produced by RIXS and are described by $\chi_c''(\textbf{q}, \omega)$. As the core level is localized, transitions can involve any $k_e$ momentum of the valence dispersing states. (b) The momentum conservation in the x-ray scattering process allows the exploration of different $(k'_e,k_e)$ combinations. (c) The RIXS spectrum for a given value of \textbf{q} is characterized by the particle-hole continuum and by sharp peaks assigned to phonon and magnon excitations; anti-Stokes transition, corresponding to an energy gain, can be present at higher temperatures. The relative intensity of excitations peaks depends on the resonance, i.e., on the core level selected for the RIXS process. In the text we refer specifically to the $2p_{3/2}$ core level of Cu and to valence states of mostly Cu $3d$ character.} 
\end{figure}

For convenience, we recall here how RIXS works for cuprates with the help of Fig. \ref{cartoon RIXS}. The resonant absorption of an x-ray photon leads to the transition of one electron from the core $2p$ to a valence $3d$ state. This highly excited configuration rapidly decays via a variety of channels. One of them implies the transition of a different $3d$ electron to fill the $2p$ hole with the emission of a photon. The excited state left behind can be a localized orbital excitation, a spin-flip (magnon) excitation, or a band-like $e$-$h$ pair.  If the very same electron decays back an elastic process takes place, or one or more phonons can be excited in the final state. In all those cases the emitted photon energy differs from the incident one by the energy of the exited state. Similarly, the momentum is conserved in the x-ray scattering process too, allowing for the determination of the energy-momentum relation by RIXS, i.e., the dispersion relation for collective excitations.

In this picture, the resonant character of RIXS is particularly beneficial, because it enhances the valence charge excitation signal more than that of phonons. Indeed, in the case of cuprates the $e$-$h$ continuum signal, growing with doping, is clearly distinguishable in RIXS spectra, although superimposed to the optical phonons in the low energy window.

\begin{figure}
\centering
\includegraphics[scale=0.5]{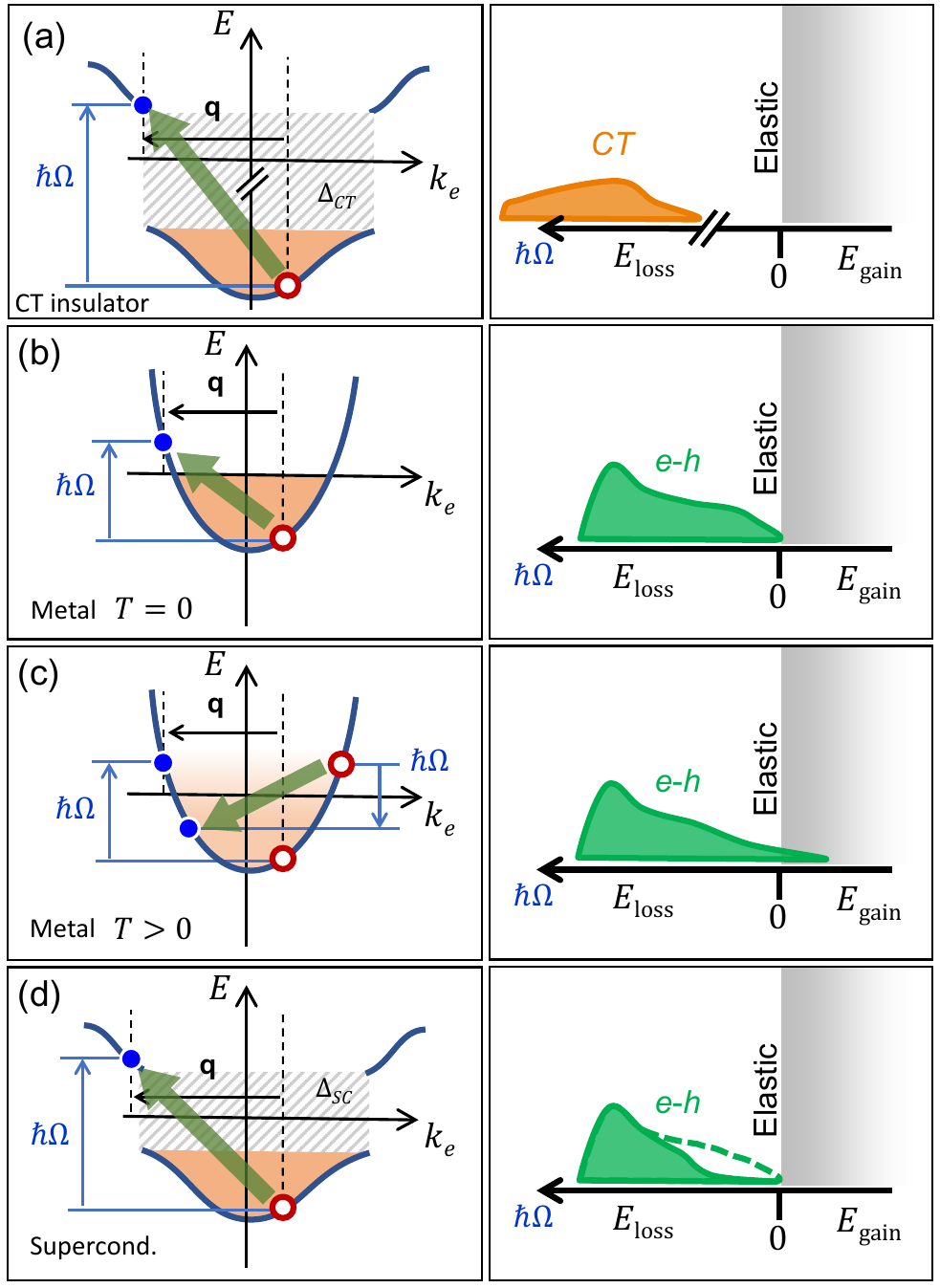}
\caption{\label{cartoon Lindhard} The particle-hole excitations and the gap opening. The schematics of the transition across the Fermi level are shown on the left, with the respective cartoon of the $e$-$h$ continuum in an energy loss spectrum on the right. In order to highlight the temperature and sample dependence of the charge excitations in a RIXS spectrum, we show the four cases of a fully gapped charge transfer insulator (a), of a metal at $T = 0$\,K (b), a metal at $T > 0$\,K (c) and a superconductor at $T = 0$\,K (d). In panel (a), the mid-IR region of the spectrum does not host any charge continuum since the system is isotropically gapped by the charge transfer (CT) energy, as opposed to the metallic and superconducting examples. This represents the case of the YBCO AF samples used for this work.}
\end{figure}

With the aim of assessing the opportunity of using RIXS to determine some of the properties of $\chi_c(\textbf{q}, \omega)$, especially the contribution of $e$-$h$ excitations, we present here a careful study of the temperature dependence of RIXS spectra of a high $T_c$ superconducting (HTS) cuprate, namely YBa$_2$Cu$_3$O$_{7-\delta}$ (YBCO). Indeed, the interplay between orbital, charge, spin and lattice degrees of freedom is particularly relevant in this class of materials \cite{{keimer2015quantum},{Scalapino},{Fradkin}}, and this complexity has so far frustrated the multiple attempts of properly explaining both the superconducting (SC) and the normal state of cuprates. Although the RIXS intensity can be connected to the dynamical charge susceptibility theoretically \cite{{Ament},{Jia},{Marra},{Schulke}}, a systematic experimental work is still missing \cite{{Koga},{Huang}}.
In the case of superconducting cuprates it is important to notice that $\chi''_c(\textbf{q}, \omega)$ should be sensitive to the opening of the superconducting gap across the Fermi level. The idea is relatively simple, as sketched in Fig. \ref{cartoon Lindhard}. Assuming that one can single out the $e$-$h$ continuum in the RIXS spectra, the low-temperature spectral function would differ from the high-temperature one for the lack of energy-gain intensity if the sample is a normal metal, and for a reduction also of the energy loss intensity if an energy gap, e.g. the superconducting gap (SG) or the pseudogap (PG), is present across the Fermi level.  Hence, RIXS can give access to the phenomenology of the SG and of the PG of HTSs. This is particularly interesting for those cuprates, e.g. ReBa$_2$Cu$_3$O$_{7-\delta}$, intrinsically incompatible with surface sensitive techniques, like angle resolved photoemission (ARPES) and scanning tunneling microscopy and spectroscopy (STM, STS).

The observation by RIXS of the SG and of the PG was enabled some years ago by the realization of instruments of adequate energy resolution, initially at the ESRF \cite{Brookes} and later at Diamond Light Source \cite{Zhou}, NSLS II \cite{Dvorak} and other synchrotrons, which must be of the order of the SG width ($\simeq40$ meV) or smaller. A pioneering work on BSCCO was published by Suzuki and collaborators in Ref. \onlinecite{Suzuki}, with a successful though partial demonstration. Indeed, it was found that superimposed to the $e$-$h$ continuum other spectral features are heavily dependent on temperature, therefore complicating the bare use of temperature differential measurements for the distillation of the pure $e$-$h$ signal form the RIXS spectra.

Building on that experience, here we propose a method to study both the SG and the PG opening for an underdoped YBa$_2$Cu$_3$O$_{7-\delta}$ (YBCO UD) ($p\simeq0.15$) by means of RIXS as a function of temperature. Complementing the experimental activity with calculations of the charge dynamical structure factor, we selected a hot-spot in the \textbf{q} space to maximize the gaps opening effects. We then performed a temperature dependence of RIXS spectra at Cu L$_3$ edge for both the YBCO UD and for an undoped insulating sample (YBCO AF), taken as a reference. Following a thorough data analysis of the RIXS temperature dependence, we are able to detect both the SG and the PG features in the low energy integrated intensity in the YBCO UD data, whereas in the YBCO AF no effect is visible, as expected. Our results demonstrate the possibility to track changes in the charge susceptibility with RIXS, opening a new route towards the study of low energy charge excitations, of the SG and of the PG in cuprates. The method can be extended to any other materials with a high quality RIXS response.

Section \ref{Methods} explains more thoroughly how the electron-hole continuum can be distinguished in a RIXS spectrum, it shows the results of our calculations of the dynamical charge structure factor and it describes the experimental conditions. Section \ref{exp_res} shows the RIXS experimental results and illustrates the different data analysis approaches. Finally in section \ref{conl_persp} the main conclusions of this work are highlighted with future perspectives.

\section{Methods}\label{Methods}
\subsection{$\chi_c(\textbf{q}, \omega)$ and the $e$-$h$ continuum in RIXS}
\begin{figure*}
\includegraphics[scale=0.8]{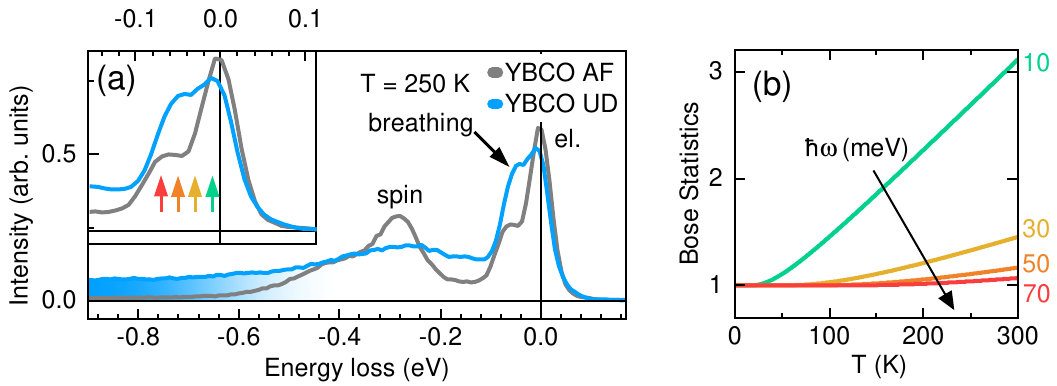}
\caption{\label{Bose and AF vs UD}(a) Example of RIXS spectra of underdoped (UD) and undoped (AF) YBCO, restricted to the energy loss window equivalent to mid-infrared optical range ($T = 250$\,K, $\textbf{q}_{\parallel} = (-0.4, 0)$). The main spectral features are the elastic peak (el), the breathing phonon (breathing) and the magnon/paramangnon (spin); the particle-hole continuum, present only in the doped sample, is highlighted with a blue shaded area. In the inset, for a narrower energy range, the arrows indicate the energy of the relative Bose curve in panel (b). (b) Bose statistics as a function of temperature for different energies labelled in meV at the right edge of the diagram.}
\end{figure*}

As mentioned above, the RIXS cross section is not proportional to $S_T(\textbf{q}, \omega)$, differently than in IXS, EELS, INS and Raman scattering. This is because in all these techniques the scattering cross section is dictated by the total charge density, which is mostly concentrated very close to the nucleus (core levels) and the valence electrons contribute marginally to the total scattering probability. Conversely, in RIXS the scattering by valence electrons is hugely amplified, whereas the scattering by the core electrons is little or at all affected by the resonance.  Consequently, in RIXS the $e$-$h$ continuum can emerge from underneath the phonon signal, which monopolizes the spectra of other scattering techniques. The two contributions have thus to be analyzed separately. Some practical issues complicate the task: the energy resolution in RIXS is still not optimal to single out in a clean way the sharp phonon peaks, the electron-phonon coupling mixes the two type of excitations and  both sets of (bosonic) excitations follow the Bose-Einstein statistics as a function of temperature. Moreover a spurious signal from defects in the bulk and at the surface of samples, mostly elastic and temperature independent, is also present with sample-dependent intensity. Therefore, the charge component of the RIXS signal can be expressed as
\begin{equation}
    I_T(\textbf{q}, \omega) = A(\textbf{q}, \omega) + S^R_T(\textbf{q}, \omega)
    \label{RIXS_assumption}
\end{equation}
where $ S^R_T(\textbf{q}, \omega)$ stands for the resonant dynamical structure factor, for which $e$-$h$ and phonons are weighted differently than in IXS, and $A(\textbf{q},\omega)$ is a temperature independent term which takes into account spurious signals. Finally, we have to mention that also orbital and spin excitations are present in RIXS spectra, which in cuprates manifest at higher energy levels compared to features associated with gap opening. The ratio between the spin and charge components of the dynamical susceptibility in a RIXS spectrum is determined by both cross-section considerations and  the intrinsic physics of the material. The experimental geometry can be tailored to favor either spin flip or  non-spin flip processes, as widely discussed theoretically \cite{Ament} and demonstrated experimentally \cite{Minola}. However, the underlying physics of the system also plays a crucial role. In Fig. \ref{Bose and AF vs UD}(a) we present a comparison between RIXS spectra of two YBCO samples: one nearly undoped and insulating (AF) and the other underdoped with a superconducting temperature $T_c=86$\,K (UD). The blue shaded area indicates the $e$-$h$ continuum, exclusively present in the UD sample. As depicted in the cartoon in Fig. \ref{cartoon Lindhard} and confirmed by the spectra in Fig. \ref{Bose and AF vs UD}(a), going from an insulating to a metallic sample enhances the charge contribution to the dynamical susceptibility in the mid-IR region of the spectrum due to the emergence of the particle-hole continuum.

Moreover, from Fig. \ref{Bose and AF vs UD}(a) We can notice that it is not evident how to disentangle the $e$-$h$ continuum from the phonon signal at relatively low energies. 
When dealing with hole-doped cuprates, one has to consider also other contributions to RIXS spectra, namely charge density waves (at zero energy loss) and charge density fluctuations (CDF) with up to few tens of meV of energy \cite{nuovoArpaia, ReviewCDF}. The latter also follow the Bose statistics but they strongly mix with both $e$-$h$ excitations and phonons close to the ordering wave-vector, where their statistical behavior is more complex \cite{nuovoArpaia}. In Fig. \ref{Bose and AF vs UD}(b) the scaling of the Bose statistics with temperature for different characteristic energies is presented. We see that intensities of all features below 30 meV, i.e., CDF, $e$-$h$ and medium and low energy phonons (Cu-O bond-bending/buckling optical and acoustic branches \cite{{BraicovichPhon},{DevereauxPhon},{PintschoviusUNO},{PintschoviusDUE},{PintschoviusTRE}}) increase with temperature. Instead the Cu-O bond-stretching/breathing modes, higher in energy ($\sim$ 65 meV), are expected to be less sensitive to temperature variations, as long as $T \leq 300$\,K  \cite{{BraicovichPhon},{DevereauxPhon},{PintschoviusUNO},{PintschoviusDUE},{PintschoviusTRE}}.

The choice of the optimum \textbf{q} where to perform the RIXS measurement so to maximize the effects of SG and PG opening is therefore crucial and not straightforward, also due to the strong anisotropy of both gaps in cuprates and to the electron-correlation effects. Ideally one should calculate $S^R_T(\textbf{q}, \omega)$ with a realistic model able to describe the electronic states, the gaps and the resonant scattering process. But this is not yet possible. It is anyway useful to look at predictions from simpler calculations. Similarly to Ref. \cite{Suzuki}, we have computed the valence-electron response function using a single Hubbard band, while keeping clearly in mind the differences between the calculated $S_T(\textbf{q},\omega)$ and a real RIXS spectrum. The results are presented in the next subsection.

\begin{figure*}
\includegraphics[scale=0.65]{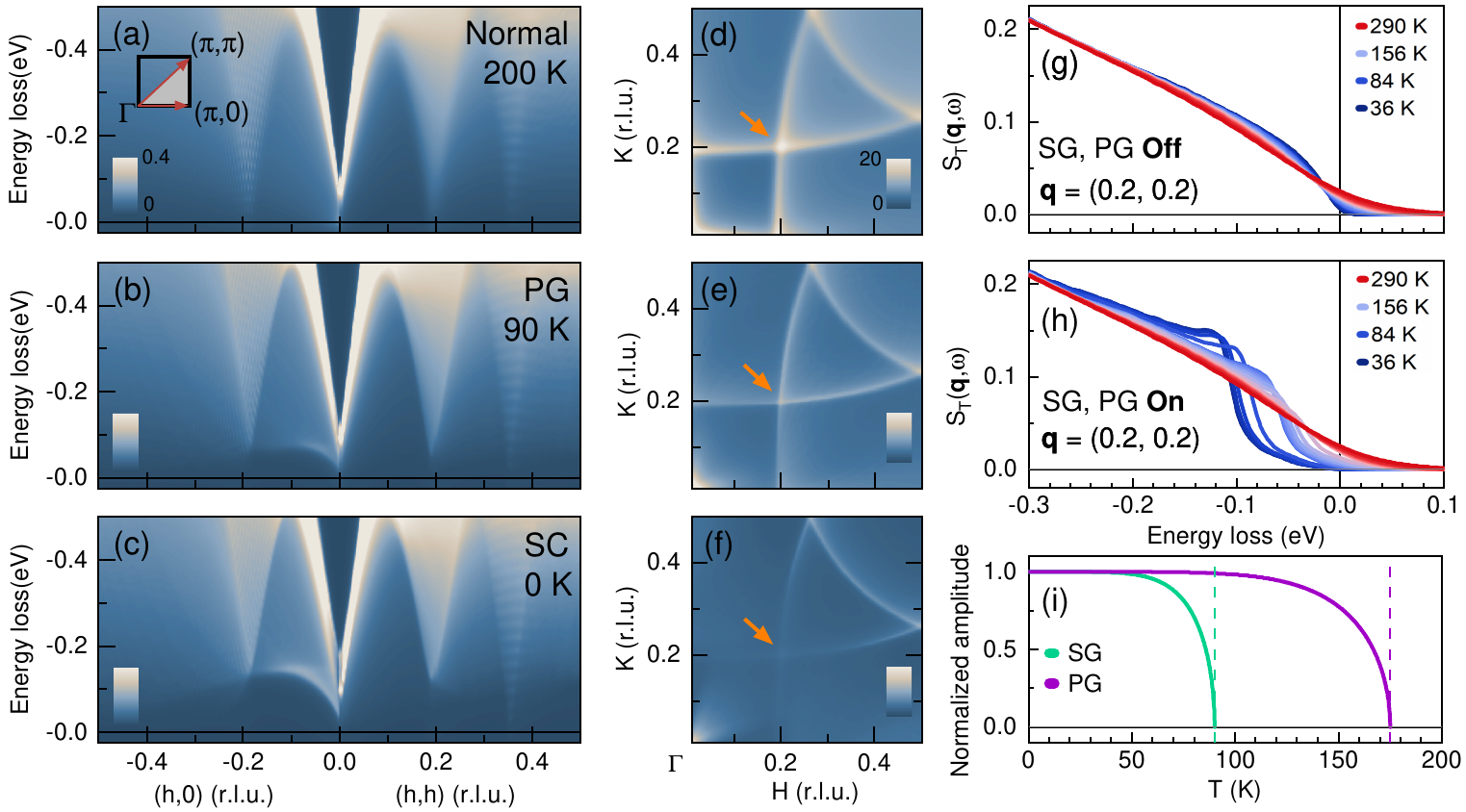}
\caption{\label{simulations} Outcomes of the dynamical structure factor calculations. (a), (b), (c) Dispersive maps along the (h,0) and (h,h) directions in the reciprocal space at $T = 200$ K, 90 K, 0 K respectively, i.e. in the normal, PG and SC phases. The intensity colour-scale is kept the same for all the different cases. (d), (e), (f) Maps in the \textbf{q} space of the integral of the simulated spectra in the first 50 meV. Again the maps are calculated respectively at 200 K, 90 K and 0 K and the intensity colour-scale is kept the same for all the different cases. (g), (h) Calculated temperature dependence spectra at \textbf{q} = (0.20, 0.20) r.l.u. from 36 K to 290 K respectively without any gap opening and with the SG and the PG. (i) Critical temperature dependence of the energy of the two gaps, normalized to their 0\,K value, as used in the calculations.}
\end{figure*}

\subsection{Calculations of the valence charge response in RIXS spectra}

In order to calculate $S_T(\textbf{q}, \omega)$ a model for $\chi_c''(\textbf{q},\omega)$ is needed. Approaches with different levels of complexity can be adopted to perform these calculations, depending on the scope of the model itself.
Here we have exploited the Lindhard function \cite{Lindhard}
\begin{equation}
    \chi_c(\textbf{q},\omega) = \sum_{k} \frac{f_{\textbf{k}-\textbf{q}} - f_\textbf{k}}{\hbar(\omega + i\Gamma) + E_{\textbf{k}-\textbf{q}} - E_\textbf{k}}
	\label{lindhard_function}
\end{equation}
where $\hbar\omega$ is the photon energy, \textbf{k} is the electron's reciprocal space vector, $f$ is the Fermi distribution, $E_\kappa$ is the band structure of the material under consideration (YBCO in our case), \textbf{q} is the momentum transfer, and $\Gamma$ takes into account the experimental broadening.
The Lindhard susceptibility describes successfully systems where the non-interacting electrons approximation is valid. Starting with a single band, it makes a convolution between occupied and empty states separated by an energy $\hbar\omega$ and a momentum $\textbf{q}$. Thus, it represents a drastically oversimplified description of cuprates, because it ignores the electron-electron correlation.
However, with the implementation of the proper band structure and of the SG and PG opening, this tool can provide a significant input for the choice of \textbf{q} in our experiment, because correlation effects are expected to move spectral weight more along the energy axis than in momentum space.
Since an experimentally measured band structure is not available for YBCO \cite{Damascelli}, we have derived it from the eight orbitals tight binding Hamiltonian from Ref. \cite{{AndersenUNO},{AndersenDUE},{AndersenTRE}}. Then, we selected for the calculations the only band crossing the Fermi level, since the Lindhard susceptibility model works under the single band approximation. 
To account for the SG and the PG we adopted the same symmetry and temperature dependence determined by ARPES for Bi2212 in Refs. \cite{{Vishik},{WSLGaps},{hashimotoGap}}. For the temperature dependence of the PG width, we adopted the identical critical behavior as observed for the SG, as shown in Fig. \ref{simulations}(i).  This choice was made for simplicity, given the absence of a more precisely defined behavior. We used amplitude values of 40 meV and 20 meV for the SG and PG respectively, referring to Bi2212 data in Ref. \onlinecite{Vishik}.
Conversely, the band structure itself remains unaltered with temperature in our calculations. The role of temperature is confined to the Fermi distribution, found in the numerator of Eq. \ref{lindhard_function}, and to the critical behaviour of the gaps.

The results of our calculations are presented in Fig. \ref{simulations}. We set $T_c$ to 90 K and $T^*$ to 175 K, and the gap value at 0\,K to be 40 \,meV. In panels (a), (b) and (c) we present the dispersive maps respectively in the normal (200 K), PG (90 K) and  SC (0 K) phases, along the (h,0) and (h,h) directions in the \textbf{q} space. As an effect of the band structure, the strongest low-energy features appear close to $\Gamma$. Upon cooling it is clear that the SG and PG openings are reflected more at certain \textbf{q} values. In order to identify these hotspots more effectively, we show the integrated intensity in the first 50 meV in the \textbf{q} space (panels (d), (e) and (f)). First, we see  that the most promising points, i.e. where the integrated intensity shows the largest differences, are also near $\Gamma$. Unfortunately, these \textbf{q} values are usually reached experimentally in near-specular geometry, where an overwhelming elastic intensity from defects at the sample surface hinders the low energy features. Therefore, we need to consider other possible solutions. The other hotspot, indicated by yellow arrows, is at $\textbf{q} \simeq (0.20, 0.20)$ r.l.u., which is the \textbf{q} value of the calculated temperature dependence spectra in panels (g) and (h).
In panel (g) we highlight the contribution of both the Fermi distribution in Eq. \ref{lindhard_function} and of the detailed balance factor \cite{Schulke} to the calculated temperature dependent $S(\textbf{q},\omega)$, switching off both the SG and the PG. Although weak, the effect is evident and represents a further temperature dependent component of the low energy scale.
In panel (h) instead we show the same calculated spectra with the gaps switched on.

\subsection{RIXS experiment}

\begin{figure}
\includegraphics[scale=0.33]{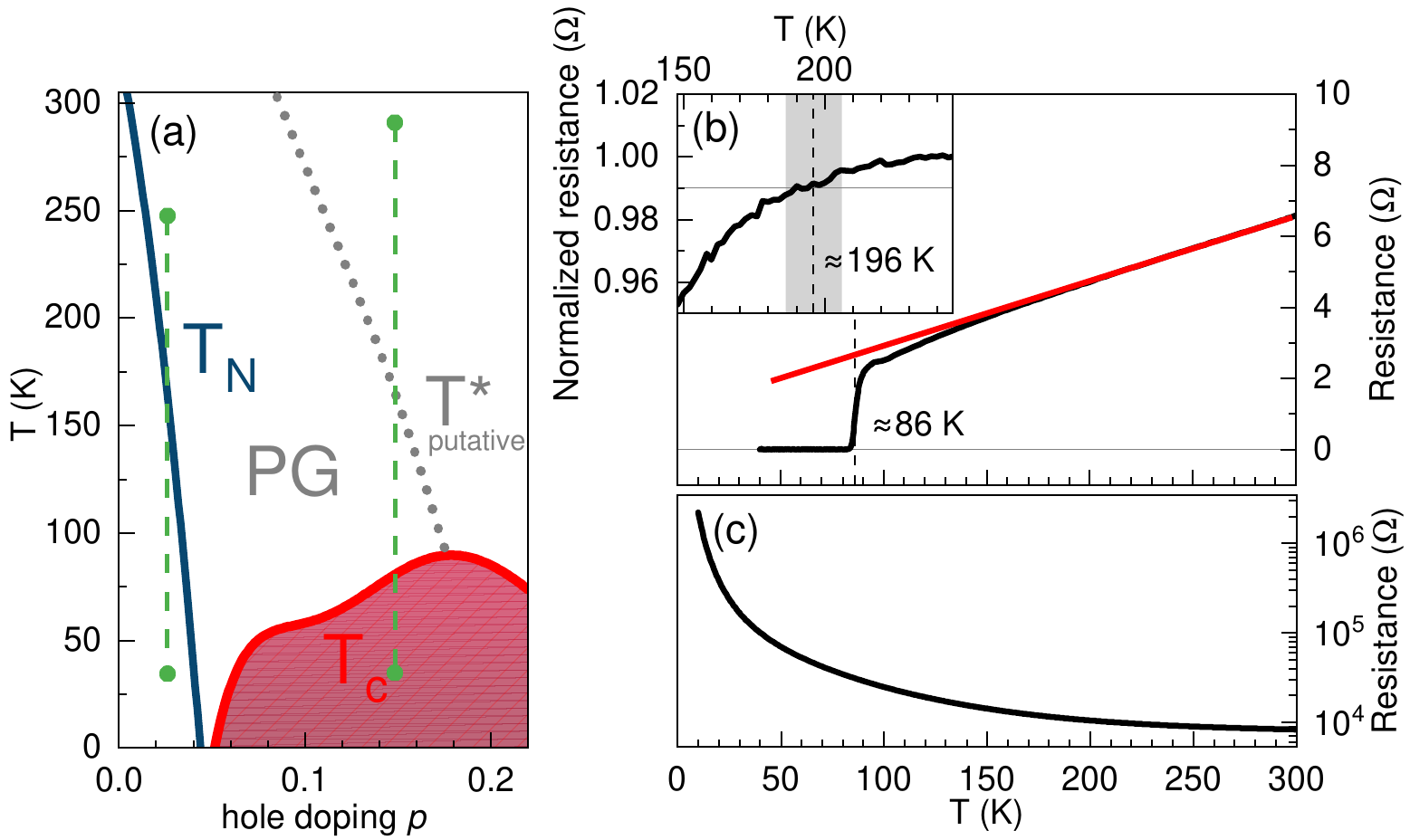}
\caption{\label{phase diagram and transport} (a) Schematic phase diagram for YBCO. The green dashed lines highlight the doping and temperature range we have investigated with the RIXS experiment. (b) Transport measurements of the YBCO UD sample. In red a linear fit of the high temperature resistance in the strange metal (SM) phase is shown. In the inset, we present the difference between $R$($T$) and the linear fit, in order to highlight the loss in linearity which defines the transition from the strange metal to the pseudogap region, conventionally associated to the pseudogap temperature $T^*$. (c) Resistance vs temperature $R$($T$) of the YBCO AF sample.}
\end{figure}

\begin{figure*}
\includegraphics[scale=0.65]{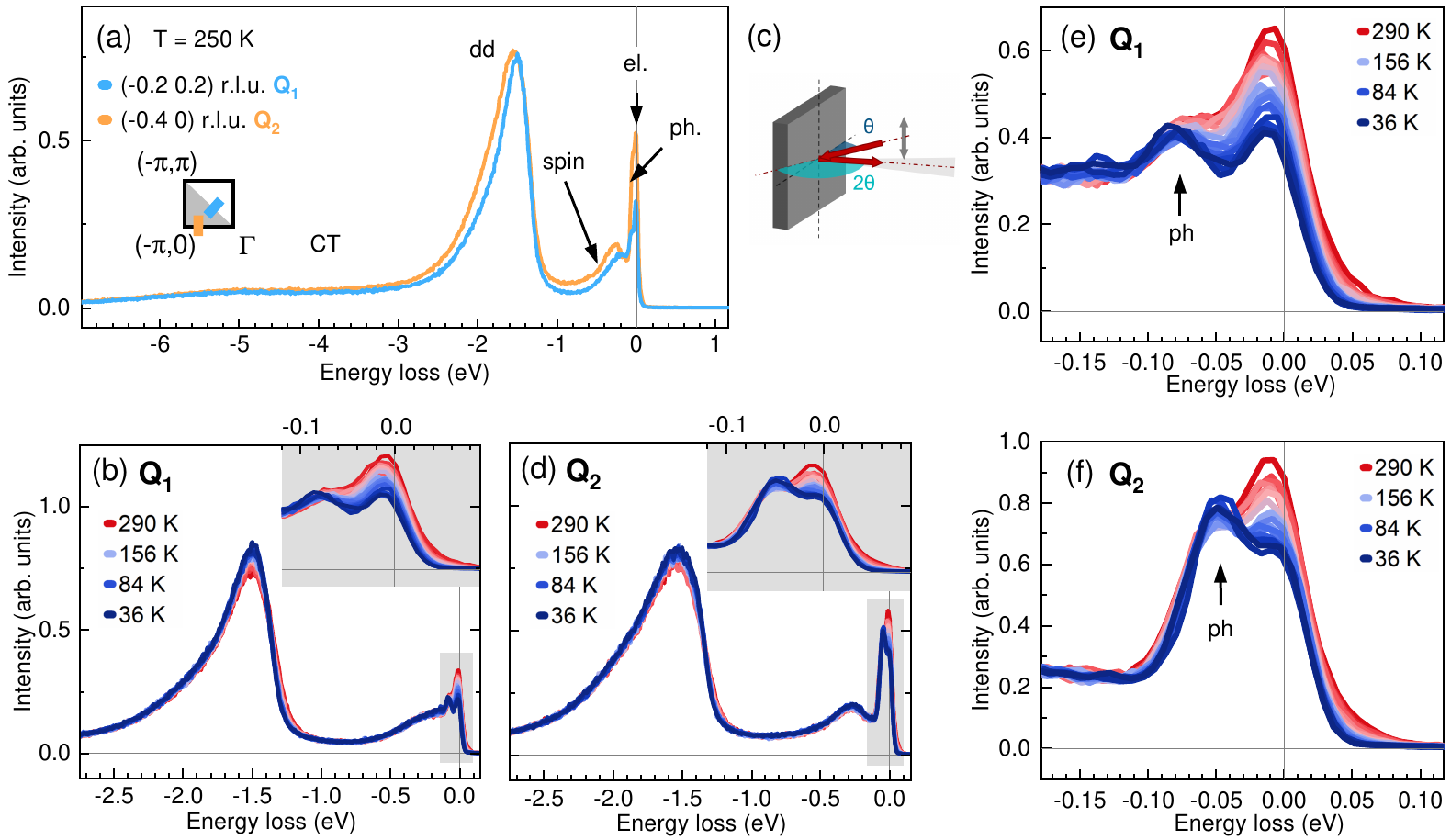}
\caption{\label{exp_description} Overview of the experimental results for YBCO UD. (a) Reference spectra for the two geometries listed in the legend, taken at $T$ = 250 K. The main spectral features are indicated by labels and the two points in the \textbf{q} space are highlighted in the inset. (b), (d) Temperature dependence of RIXS spectra respectively at (-0.2, 0.2) r.l.u. (\textbf{Q$_1$}) and (-0.4, 0) r.l.u. (\textbf{Q$_2$}). The temperatures go from 290 K (red) to 36 K (blue). In the insets, the low energy scale is highlighted. The spectra are normalized to the $I_0$. (c) Cartoon of the scattering geometry. The incoming and outgoing beams are represented by red arrows. The incident angle ($\theta$) is depicted in blue and the scattering angle ($2\theta$) in light-blue. The grey arrow represents the polarization of the incoming beam. (e), (f) Low energy spectral region at \textbf{Q$_1$} and \textbf{Q$_2$}. The spectra are normalized to the integrated intensity of the $dd$ excitations (from -3 meV to -1 meV). The breathing mode phonon peak (ph) is highlighted in both the panels.}
\end{figure*}

\begin{figure*}
\includegraphics[scale=0.65]{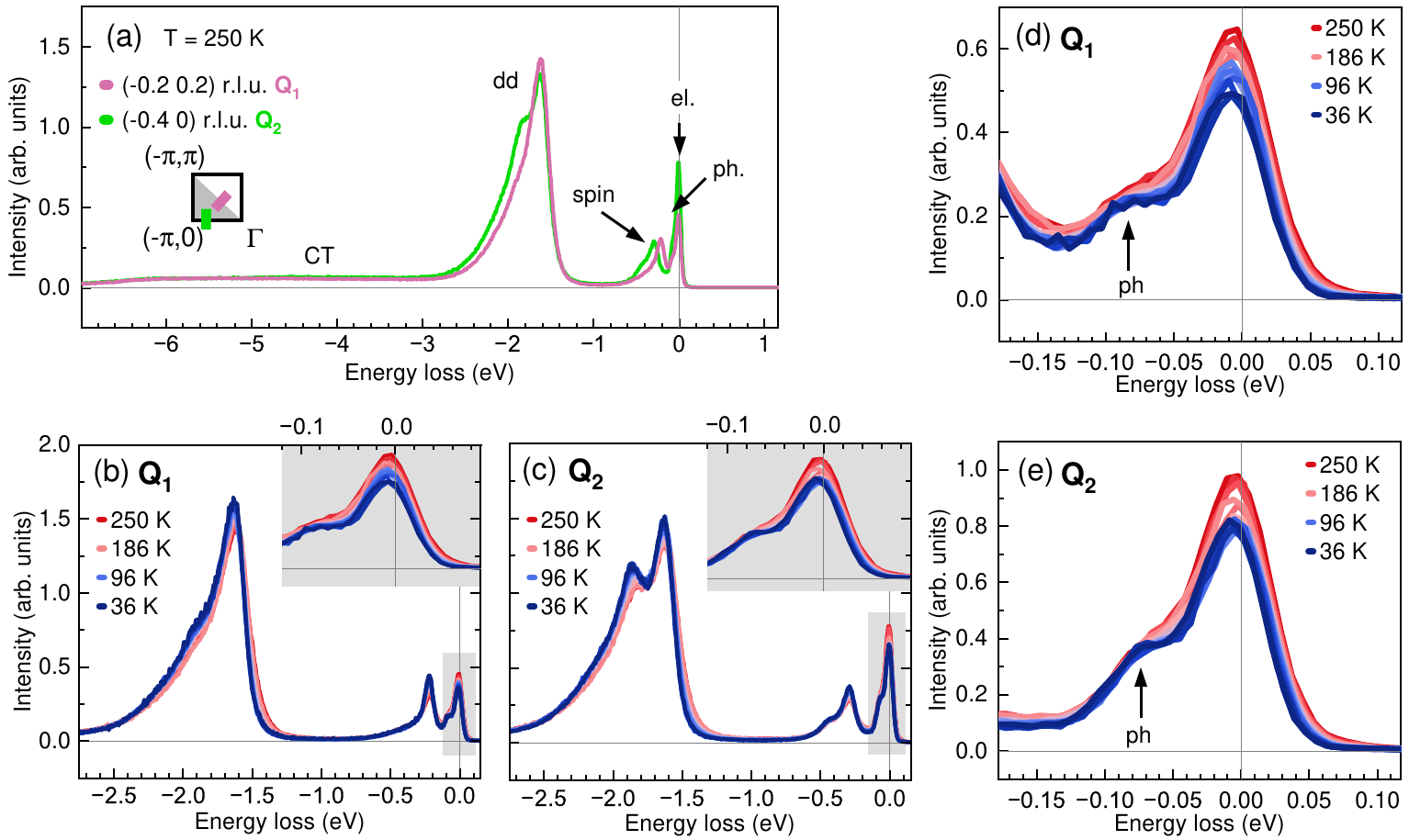}
\caption{\label{exp_descriptionAF} Overview of the experimental results for YBCO AF. (a) Reference spectra for the two geometries listed in the legend, taken at $T$ = 250 K. The main spectral features are indicated by labels and the two points in the \textbf{q} space are highlighted in the inset. (b), (c) Temperature dependence of RIXS spectra respectively at (-0.2, 0.2) r.l.u. (\textbf{Q$_1$}) and (-0.4, 0) r.l.u. (\textbf{Q$_2$}). The temperatures go from 250 K (red) to 36 K (blue). In the insets, the low energy scale is highlighted. The spectra are normalized to the $I_0$. (d), (e) Low energy spectral region at \textbf{Q$_1$} and \textbf{Q$_2$}. The spectra are normalized to the integrated intensity of the $dd$ excitations (from -3 meV to -1 meV). The breathing mode phonon peak (ph) is highlighted in both the panels.}
\end{figure*}

\begin{figure}
\includegraphics[scale=0.95]{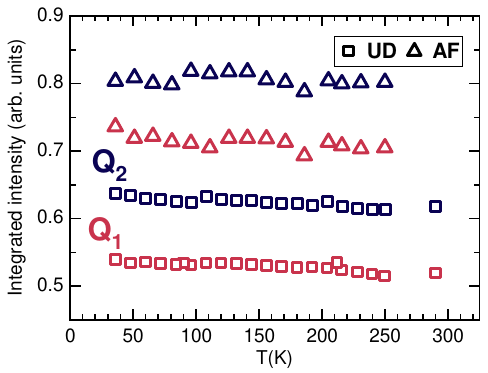}
\caption{\label{dd_intensity} Intensity of $dd$ excitations integrated from -3 eV to -1 eV for both \textbf{Q$_1$} and \textbf{Q$_2$} and for the two samples.}
\end{figure}

We have measured the temperature dependence of RIXS spectra at the Cu L$_3$ edge on two YBCO samples, the former slightly underdoped (UD) and the latter insulating (AF), in the temperature range depicted by the two green dashed lines in \ref{phase diagram and transport}(a).
The YBCO films, with thickness $t$=100 nm, have been deposited by pulsed laser deposition on 5×5 mm$^2$ (001) oriented SrTiO$_3$ substrate. Details of the growth procedure can be found in Ref. \cite{ArpaiaGrowth}. The slightly underdoped sample has been obtained using a post-annealing oxygen pressure $p_{\mathrm{ann}}= 1.5\times10^{-3}$ torr after the deposition. Here, the critical temperature (temperature where the first derivative of the $R$($T$) characteristic is maximum) is $T_c$=86 K and the PG temperature (temperature where the resistance deviates by 1\% from the linear fit) $T$*=196 K (Fig. \ref{phase diagram and transport}(b)). The doping level $p$ = 0.15 has been determined by the combined knowledge of $T_c$ and of the $c$-axis length \cite{ArpaiaGrowth}. In the YBCO AF case, the sample has undergone a post-annealing with an oxygen pressure of $p_{\mathrm{ann}}= 1\times10^{-5}$ torr, which translates into a doping level $p<0.04$.

RIXS spectra were measured at the ID32 beamline of the European Synchrotron (ESRF) in Grenoble, France \cite{Brookes}. We worked in the high efficiency configuration of both the monochromator and the beamline to maximise the incoming photon flux. The energy resolution ($\simeq40$\,meV for the YBCO UD measurements and $\simeq50$\,meV for the YBCO AF) was determined from a pure elastic line taken at specular condition. The momentum resolution is defined by the angular acceptance of the spectrometer in the scattering plane (20 mrad) \cite{Brookes} and can be estimated to be $\simeq0.01$ r.l.u..
A sketch of the scattering geometry is shown in Fig. \ref{exp_description}(c). We employed $\sigma$ polarization of the incoming x-ray photons, scattering angle $2\theta = 149.5^\circ$ to maximize the momentum transfer \textbf{q}, and grazing-incidence configuration, corresponding to negative $\textbf{q}_{\parallel}$ in our convention.

We selected two \textbf{q} points following the results of our calculations. \textbf{Q$_1$}, in $(-0.2, 0.2)$ r.l.u., is expected to be a hotspot for observing large effects due to the SG and PG opening.  On the contrary, \textbf{Q$_2$} in $(-0.4, 0)$ r.l.u. serves as a reference for the other temperature dependent features in the low energy spectral region because, according both to our calculations and to the previous measurements \cite{Suzuki}, the SG and to the PG should influence very little the spectra in this point.

As for YBCO UD, we acquired spectra in parallel at the two projections of wave vector while scanning the temperatures from 290\,K down to 36\,K in steps of about 12\,K. Each spectrum took one hour of acquisition, with the only exception of the one at 250 K, exploited as a reference in the analysis, which has been measured for two hours.
In Fig. \ref{exp_description}(a) wide spectra at $T = 250$ K are presented for the two geometries. The two \textbf{q} points are indicated in the inset.

In order to strengthen our results, we performed the same temperature dependence on the YBCO AF, whose temperature evolution should be determined exclusively by low energy phonons, as explained in section \ref{Methods}A. We adopted the very same experimental geometry and we followed the experimental protocol of the UD sample. An overview of the results is shown in Fig. \ref{exp_descriptionAF}. In particular, in panel (a) a comparison of the spectra in the two geometries for the AF case is presented.

\section{Experimental results}\label{exp_res}
\subsection{RIXS temperature dependence}
\begin{figure*}
\includegraphics[scale=0.65]{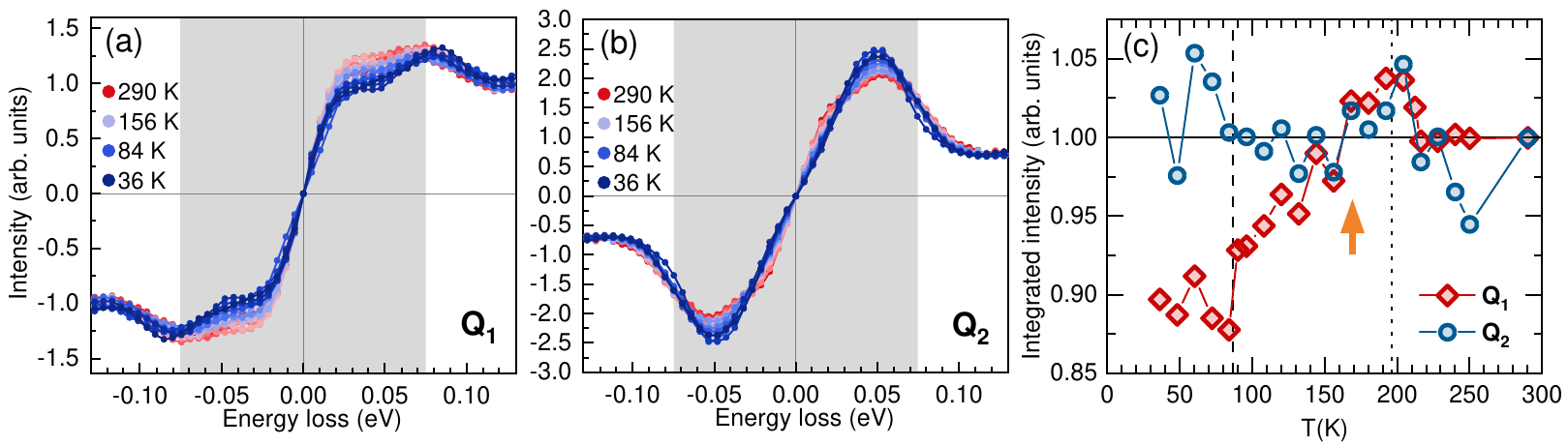}
\caption{\label{antisimm}(a) Low energy antisymmetrized temperature dependent spectra for \textbf{Q$_1$}. (b) Low energy antisymmetrized temperature dependent spectra for \textbf{Q$_2$}. (c) Integrated intensity of the absolute value of (a) and (b) for the grey shaded areas, normalized to the 290 K point. The error bars are within the symbol size. The dashed and dotted lines indicate respectively $T_c$ and $T^*$ extracted from transport measurements, as illustrated in Fig. \ref{phase diagram and transport}.} 
\end{figure*}

In Figs. \ref{exp_description}(b), \ref{exp_description}(d), \ref{exp_descriptionAF}(b) and \ref{exp_descriptionAF}(c) we show an overview of the temperature dependent RIXS spectra normalized to the $I_0$, for both the geometries and for the two samples. As for the UD case, at both momenta, the low energy region shows a marked temperature dependence. The $dd$ excitations broad peak is the other spectral feature that exhibits a clear temperature dependence, although weaker, consisting in a softening and a broadening of the peak with temperature. Although the experimental conditions do not favour the spin excitations, it seems clear that the paramagnon is mostly insensitive to temperature.
In the AF case the temperature effects are more evident in the $dd$ excitations, again with a softening at larger temperatures. Although weaker, the quasi-elastic region is still showing a very clear trend, which can be assigned to low energy phonons, as discussed later.

In Figs. \ref{exp_description}(e), \ref{exp_description}(f), \ref{exp_descriptionAF}(d) and \ref{exp_descriptionAF}(e) we focus instead on the low energy region. In order to clear out residual changes in intensity due to beamline instabilities, these spectra are normalized to the $dd$ peak integral, from -3 eV to -1 eV, with the reasonable assumption that the overall intensity of the peak should remain almost constant at different temperatures. Indeed, as shown in Fig. \ref{dd_intensity}, the variation of the total $dd$ intensity with temperature is always very small, and unaffected by the SC transition in the UD case (squares). To further confirm the monotonous trend, in Fig. \ref{dd_intensity} the same integrated intensities in both geometries for the undoped (AF) YBCO sample are shown (triangles).
This internal normalization is useful since we are looking for small intensity modulations due to the SG and PG opening. Comparing Fig. \ref{exp_description} panels (e) and (f) with the insets of panels (b) and (d) it is clear that this different normalization approach does not significantly change the overall results.

In Figs. \ref{exp_description}(e), \ref{exp_description}(f), \ref{exp_descriptionAF}(d) and \ref{exp_descriptionAF}(e) the change in intensity within the low energy window is highlighted. As expected, for both \textbf{q} values the bond-stretching phonon is almost temperature independent and, conversely, at lower energies across the Fermi level spectra intensity grows monotonically with $T$. These results confirm that low-energy phonons and CDF (in the UD case) are indeed contributing to the temperature dependence in the low energy scale. In particular, the clear temperature evolution of the quasi-elastic intensity also in the AF case further confirms the role of low energy phonons in the temperature dependence of this spectral region, since  low-energy charge contributions are almost entirely suppressed at this doping level.
It is thus clear that the observation of the SG and PG features requires cleaning the signal from these other temperature dependent excitations, which hinder the direct observation of gap-related information in the raw spectra.

\subsection{Low energy scale intensity evolution}\label{antisimmet}
As a first approach to better monitor the temperature dependence of the low energy spectral weight, we adopt the well established antisymmetrization method to retrieve $\chi''_c(\textbf{q}, \omega)$ from the spectra, as illustrated in Ref. \cite{Vig}.
In order to obtain an estimate for $\chi''_c(\textbf{q}, \omega)$ we need to remove the Bose factor that multiplies $\chi''_c(\textbf{q}, \omega)$ in the fluctuation dissipation theorem \cite{Schulke} as follows:
\begin{equation}
    \chi''_c(\textbf{q}, \omega)=-\pi\left[I_T(\textbf{q},\omega)-I_T(\textbf{q},-\omega)\right]
\end{equation}
where $I_T(\textbf{q},\omega)$ is defined in Eq. \ref{RIXS_assumption}. As explained in Ref. \cite{Vig}, the advantage of this approach is to cancel out the Bose factor contribution by automatically accounting for it in the the antisymmetrization of the experimental data with respect to the zero energy point, exploiting the odd nature of $\chi''_c(\textbf{q}, \omega)$. Ideally, any spectral contribution following the Bose statistics would result in a temperature independent behaviour, therefore highlighting the discrepant spectral features as special. We note that the spurious $A(\textbf{q}, \omega)$ term gets also cancelled out if it is of pure elastic scattering, i.e., $\omega = 0$ so that  $A(\textbf{q}, \omega) = A(\textbf{q}, -\omega)$ for any \textbf{q}.
Figs. \ref{antisimm}(a) and (b) show $\chi''_c(\textbf{q}, \omega)$ for both geometries.
In order to catch any signature of the SG and the PG, we take the absolute value of $\chi''_c(\textbf{q}, \omega)$ and we integrate it within $\pm75$ meV. The result of the integration is shown in Fig. \ref{antisimm}(c).
The difference between \textbf{Q$_1$} and \textbf{Q$_2$} is evident. Within the experimental error, the \textbf{Q$_2$} intensity remains unperturbed, meaning that the opening of both the PG and the SG does not affect the spectral intensity in the low energy range, as predicted by our calculations. Although some temperature effects are still visible in the first 30 meV (see Fig. \ref{antisimm}(b)), this is compensated by higher energy modulation and thus not reflected in the integrated intensity of Fig. \ref{antisimm}(c).
On the other hand, the temperature evolution of \textbf{Q$_1$} is more structured. Starting from high temperatures down to $\simeq170$ K (orange arrow in Fig. \ref{antisimm}(c)) the data points are superimposed to \textbf{Q$_2$}. Then across the putative PG opening temperature, the \textbf{Q$_1$} integrated intensity deviates from that of \textbf{Q$_2$}. However, we note that this depletion starts at slightly lower temperatures with respect to what is extracted from transport measurements (dotted line in Fig. \ref{antisimm}(c)).
The second turning point corresponds with T$_c$ measured in transport, thus we attribute it to the SG opening.

\subsection{Influence of low energy phonons and CDF}\label{Bose_analysis}

\begin{figure*}
\includegraphics[scale=0.65]{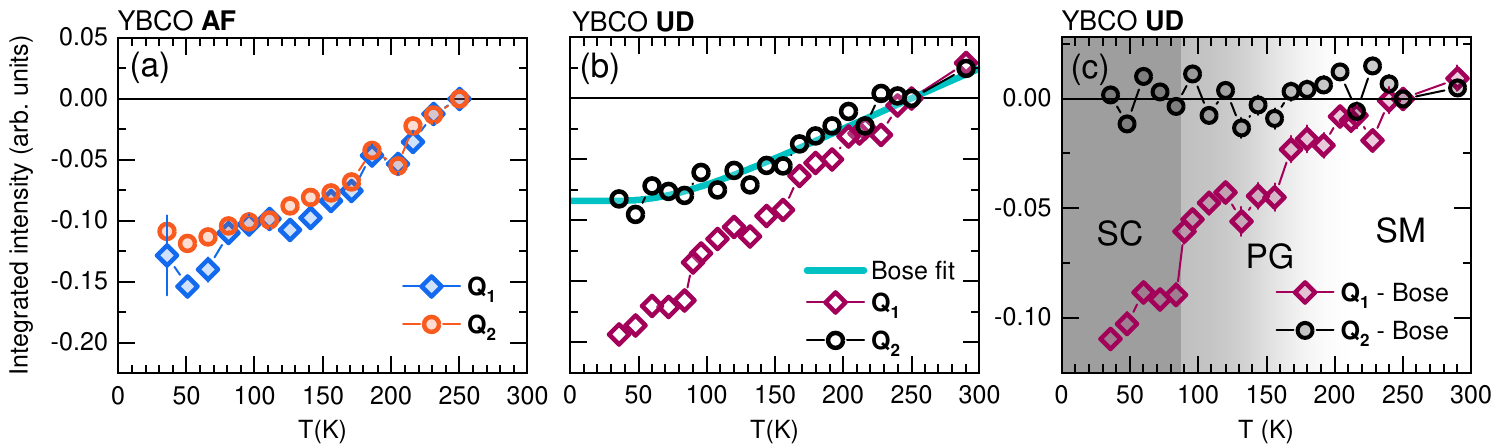}
\caption{\label{integrals} (a), (b) Integrated intensity of the difference of the spectra taken at different temperatures with respect to the one at 250 K, respectively for YBCO AF and UD and for both the geometries as indicated in the legend. The integral is stopped at 75 meV, which corresponds to the energy of the breathing phonon peak at \textbf{Q$_1$}. For the (b) panel, in light blue the fit of the \textbf{Q$_2$} with a Bose-like curve.} (c) Integrated intensity of the differences shown in panel (b) after the subtraction of the Bose fit for \textbf{Q$_2$} at both \textbf{q} values. The error bars are indicated and mostly within the symbol size. The different grey areas represent the superconducting (SC), pseudogap (PG) and strange metal (SM) phases.
\end{figure*}

\begin{figure}
\includegraphics[scale=0.70]{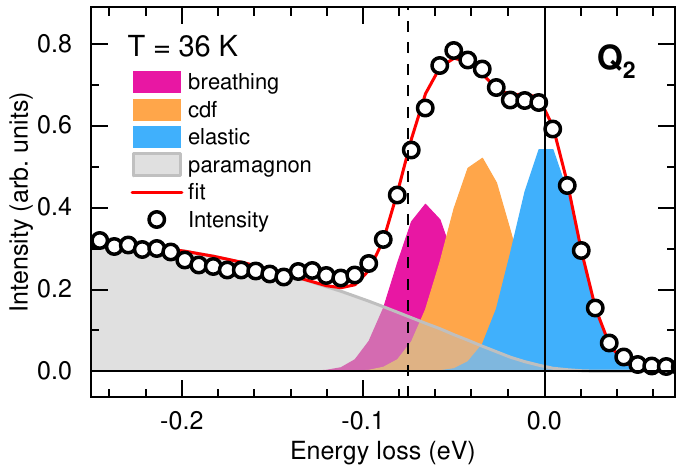}
\caption{\label{fitting_Q2} Fitting of the low energy spectral region for \textbf{Q$_2$} at 36 K. The CDF peak is reported in orange, while the elastic and breathing mode peaks are shown respectively in light blue and purple. The tail of the paramagnon feature is fitted with a damped oscillator, reported in gray. The total fit is represented by the red curve.} 
\end{figure}

To confirm these results, we analyze the temperature dependence of the low energy region from a different perspective.
Starting again from the RIXS spectra in Fig. \ref{exp_description} and Fig. \ref{exp_descriptionAF}, to amplify the temperature effects we take the differences between all the spectra and the reference one at $T$ = 250 K. Then, we calculate the integral of these differences from -75 meV, which corresponds to the energy of the breathing phonon peak at \textbf{Q$_1$}, up to +100\,meV, to include all energy gain contributions (ideally one should integrate up to $+\infty$). In order to make the two data-sets at different \textbf{q} points comparable, we normalize these values to the integrals of the sum of the different spectra and the reference, as follows:
\begin{equation}
    I(T) = \frac{\int_{-75\textrm{ meV}}^{+100\textrm{ meV}} [S(T)-S(T=250\textrm{ K})] dE}{\int_{-75\textrm{ meV}}^{+100\textrm{ meV}} [S(T)+S(T=250\textrm{ K})] dE}
    \label{integration}
\end{equation}
where $S$ here is the RIXS spectral intensity.
The results are presented in Fig. \ref{integrals}(a) and (b). Starting with the AF case (Fig. \ref{integrals}(a)) as a reference, it is evident that, within the experimental error, the trends at the two different \textbf{q} points align. 
Indeed, the main contribution to the intensity comes from low energy phonons, e.g., buckling mode and acoustic branches, which are almost non-dispersing \cite{DevereauxPhon}.
However, we cannot entirely rule out a potential quasi-isotropic contribution from charge ordering, whose evidence has been recently hinted at doping levels below $p\sim0.05$ \cite{zhao_undoped_charge}.

As opposed to the AF data, in the UD case the two geometries present clearly different trends. At high temperature, where the sample is in the normal state (Fig. \ref{phase diagram and transport}), the two integrals scale quite consistently. The first weak deviation of the \textbf{Q$_1$} data from the trend is visible below 200 K, and a more considerable intensity drop is between 90 K and 84 K. On the contrary, within the experimental uncertainties there are no jumps around $T_c$ (Fig. \ref{phase diagram and transport}(b)) for \textbf{Q$_2$}, as predicted by our calculations.
We note that the temperature evolution of the data taken at \textbf{Q$_2$} in the UD is essentially the same as that of the AF case for both \textbf{q} points.

As explained in Ref. \cite{nuovoArpaia}, the quasi-elastic temperature dependence due to CDF and low energy phonons can be well described as
\begin{equation}
    B_{\omega,I_0}(T) = A(\textbf{q}) + I_0\left[1+\frac{2}{\text{exp}(\frac{\hbar\omega}{k_BT})-1}\right]
    \label{Bose CDF}
\end{equation}
where $A(\textbf{q})$ is an additive temperature independent component, $I_0$ is the CDF plus phonons intensity at 0 K, $\omega$ is the characteristic energy of the excitation and $k_B$ is the Boltzmann constant.

In order to disentangle the SG and PG contributions to the total temperature dependence from the others (CDF, low energy phonons) for \textbf{Q$_1$}, we fit the \textbf{Q$_2$} case in Fig. \ref{integrals}(a) according to Eq. \ref{Bose CDF} and we subtract the result from the \textbf{Q$_1$} data-points.
Since both \textbf{Q$_1$} and \textbf{Q$_2$} are far from the CDF critical wave vector \textbf{q} \cite{{ArpaiaUNO},{ReviewCDF}}, the CDF temperature dependence can be considered to be the same in the two cases \cite{nuovoArpaia}.

Following the approach of Ref. \cite{nuovoArpaia}, we extract the $A(\textbf{q})$ from the spectrum at low temperature. We fit the low energy scale with 4 peaks: a damped harmonic oscillator for the paramagnon \cite{PengPRB}, two resolution-limited Gaussian peaks for the breathing phonon and the elastic scattering, and a broader Gaussian accounting for CDF and low energy phonons, adopting the fitting procedure of Ref. \cite{nuovoArpaia}\cite{BraicovichPhon}\cite{Rossi}\cite{DJHuang}. An example of the fitting is shown in Fig. \ref{fitting_Q2}.
In order to estimate $A(\textbf{q})$, we take the total intensities of the elastic and breathing phonon peaks obtained from the fitting of the spectrum at low temperature. We consider these two features as the relevant temperature independent components for the low energy scale. Then, exploiting the obtained $A(\textbf{q})$, we fit the \textbf{Q$_2$} points in Fig. \ref{integrals}(a) with Eq. \ref{Bose CDF}, considering $I_0$ and $\omega$ as free parameters, taking into account the procedure explained in Eq. \ref{integration}.
The result of this fitting is represented by the light blue curve in Fig. \ref{integrals}(a). The scaling with temperature of the low energy spectral intensity at \textbf{Q$_2$} is well reproduced by the fitting, confirming the Bosonic nature of the temperature dependence in \textbf{Q$_2$}, which can therefore be exploited to minimize the CDF and phonons contributions for \textbf{Q$_1$}.

In Fig. \ref{integrals}(b) we show the integrals after the subtraction of the same bare Bosonic intensity fit for the two geometries. For \textbf{Q$_1$} the intensity step at 87 K noticed in Fig. \ref{integrals}(a) is now more evident. We consider it as the signature of the SG opening around $T_c\simeq87\pm3$ K), determined by transport measurements in Fig. \ref{phase diagram and transport}(b).

Above $T_c$ an overall spectral weight depletion is evident below $\simeq 200$ K and can be mainly assigned to the PG opening, confirming what we obtained with the antisymmetrization method in section \ref{antisimmet}. This result is in fairly good agreement with the $T^*$, indirectly extracted from transport measurements (Fig. \ref{phase diagram and transport}(b)).
However, although the SG transition clearly shows a signature in the integrated intensity, the results in Fig. \ref{integrals}(b) do not show any sharp feature for the PG, within the experimental error.

\section{Conclusions and perspectives} \label{conl_persp}
We have measured the temperature dependent RIXS spectra at two distinct \textbf{q} points, focusing on both an underdoped YBCO sample and an undoped reference sample. Guided by Lindhard susceptibility calculations, we selected \textbf{Q$_1$} and \textbf{Q$_2$} to respectively maximize and minimize the effects on the spectra of the presence of superconducting gap and pseudogap. In consideration of the experimental difficulties (high energy resolution and high stability over long period of time to span a wide temperature range in fine steps), we analyzed the data with two different approaches.
In both cases, we have been able to decouple the electron-hole continuum temperature dependence from the other Bosonic features (phonons, CDF) and to observe both the signature of the SG and the PG in the underdoped system. The SG opening reflects into a sharp step in the spectral weight at $T_c\simeq87\pm3$ K, in good agreement with transport measurement.
Instead, the PG signature results in a smoother spectral intensity evolution above $T_c$, which extends up to $\simeq200$ K, around $T^*$ as obtained from transport measurements.
Thus, while for the SG we can extract a well defined $T_c$ value, in the case of the PG no sharp onset temperature can be identified, in agreement with other transport and spectroscopic investigations of the phenomenon.
Although our approach undergoes complications from the experimental side, we have demonstrated the sensitivity of RIXS to both the SG and the PG. This result strengthen the possibility to have access to the charge response function by means of RIXS.

Our results are particularly relevant for the physics of cuprates. 
Indeed, the longstanding debates about the pseudogap and its nature, in relation to the superconductivity (does it compete with the SC state or does it represent a sort of precursor SC phase \cite{Norman}?), and to the strange metal phase (is it a mere crossover line \cite{seibold} or does it represent the onset of a proper critical phase transition?\cite{Rev_Taillefer}) have not been entirely settled to these days. In the last years the almost accepted belief has been to derive $T^*$ from transport measurements as the onset temperature for the loss of linearity in the resistivity, which represents the main signature of the strange metal phase. Thus, transport measurements have been deeply exploited to study the PG to strange metal transition, inspiring the proposal of a quantum critical point (QCP)-based phase diagram \cite{{Taillefer},{Hussey}}. However, as shown in \ref{phase diagram and transport}(b), the $T^*$ extraction from transport measurements is indirect and the uncertainty is not negligible. 
Instead, recent ARPES measurements suggest to adopt a radically different interpretation on the role of $T^*$ \cite{Sudichen}. According to this picture, the PG would be subordinated to the strange metal phase and the critical doping level $p_c$ would represent a ``temperature independent boundary” between the strange metal and the Fermi liquid regimes, thus excluding the existence of any QCP.
Unfortunately, ARPES can be used only for few families of HTS. Hence, another spectroscopic technique assessing the PG nature is very welcome because it can help reconciling spectroscopy and transport measurements, heading for a more consistent interpretation of this puzzling regime.

\appendix
\section{Assignment of the zero of the energy}

\begin{figure*}
\includegraphics[scale=0.7]{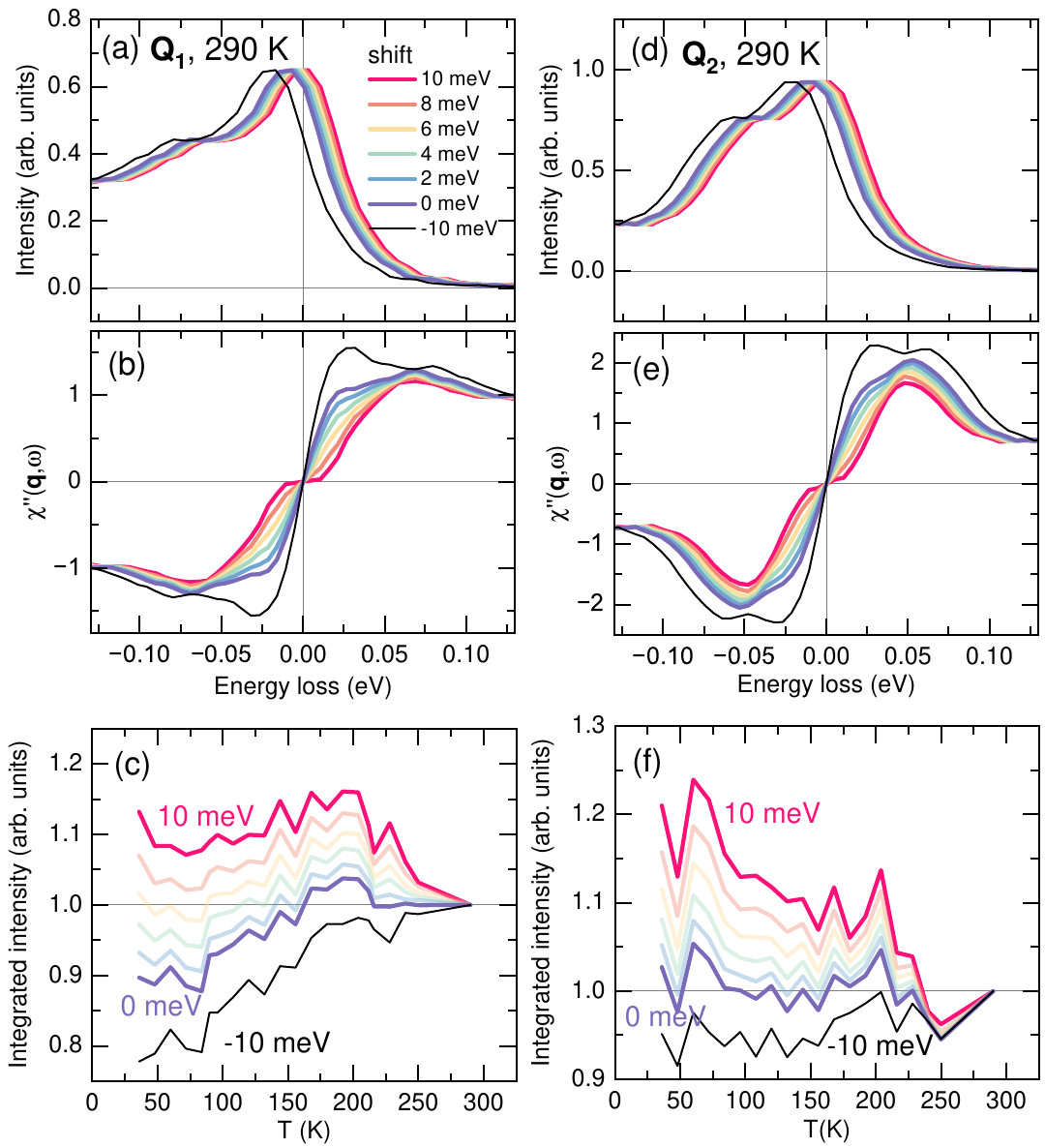}
\caption{\label{appendix_A} Antisymmetrization as a function of the energy shift, from our zero of the energy towards a zero defined by the maximum of the first peak. (a) RIXS spectra at \textbf{Q$_1$} taken at 290 K as a function of the energy shift. The colour code is the same for all the panels. (b) As in Fig. \ref{antisimm}(a,b) $\chi''_c(\textbf{q}, \omega)$ at \textbf{Q$_1$} as a function of the energy shift. Going towards positive shifts $\chi''_c(\textbf{q}, \omega)$ becomes more and more flat around zero. (c) As in Fig. \ref{antisimm}(c), normalized T dependent integrals of $\chi''_c(\textbf{q}, \omega)$ at \textbf{Q$_1$} for the different energy shifts. We highlight the integrals corresponding to our zero of the energy (0 meV), to the maximum of the first peak (10 meV) and to a shift of 10 meV in the opposite direction (-10 meV). (d), (e), (f) same as (a), (b), (c), but at Q2.}
\end{figure*}

The definition of the zero of the energy is highly debated in the community since the advent of extremely high resolution spectrometers combined with high quality samples with very weak  diffuse elastic response, whose intensity can be smaller than that of phonons at energy losses of few tens of meV of energy loss, potentially leading to erroneous assignment of zero energy elastic peaks. 
As for the present work, the antisymmetrization method proposed in section \ref{antisimm} is intrinsically affected by the zero energy location, which in fact makes it a very sensitive way of checking the quality of our assignment of the zero energy. As explained in section \ref{antisimmet}, the antisymmetrization method exploits the anti-Stokes spectral weight to cancel out the Bosonic contribution from the Stokes spectral weight, together with the temperature independent intensity.
According to the fluctuation-dissipation theorem, this results in the pure imaginary part of the response function ($\chi''_c(\textbf{q}, \omega)$).
One of the most important features of $\chi''_c(\textbf{q}, \omega)$ is that it is an odd function of the energy for any value of q and it is supposed to be asymptotically linear in energy across zero energy \cite{Schulke}. One possible test to verify the accuracy of the zero energy assignment is that of checking for the linearity range across zero of the extracted $\chi''_c(\textbf{q}, \omega)$.
In order to confirm the correctness of our zero energy definition, we repeated the analysis presented in Fig. \ref{antisimm} by shifting the spectra in steps of 2 meV up to 10 meV (where the zero of the energy corresponds to the maximum of the first peak). In figure \ref{appendix_A}(a) and (d) we show the 290 K spectra as a function of the energy shift, respectively for \textbf{Q$_1$} and \textbf{Q$_2$}. We considered the spectra at 290 K because no electronic gaps are present at such high temperatures. In Fig. \ref{appendix_A}(b) and (e) we show the resulting $\chi''_c(\textbf{q}, \omega)$ at 290 K. It is clear that while our original zero of the energy gives a $\chi''_c(\textbf{q}, \omega)$ which is linear at low energies, a departure in shape from the expected behavior occurs when shifting the spectra by 10 meV or less. The reason for this can be attributed to an overestimated anti-Stokes spectral weight that ends up cancelling the Stokes intensity upon antisymmetrization. Consequently the integration of the $\chi''_c(\textbf{q}, \omega)$ as a function of temperature for the different values of the energy shifts cannot lead to the same result, as shown in figure 2(c) and (f). In particular, in the case of \textbf{Q$_2$} (Fig. \ref{appendix_A}(f)) with our assignment of the zero of energy we find no temperature dependence in agreement with the results obtained by the very different analysis explained in section \ref{Bose_analysis}. Conversely, when changing the zero of the energy towards 10 meV the low-energy intensity integral shows a clear decrease with temperature, an effect very difficult to explain. Finally, we would like to stress that the more we shift the spectrum towards the energy gain side (positive energy shift), the more we subtract in the antisymmetrization. By going in the opposite direction (thin black lines in figure 2, negative shift), the effect of the antisymmetrization procedure on the total spectral intensity is less and less relevant and the result of the analysis tends to be closer to a normal integration of the spectra in the low energy range. This consideration suggests that a systematic error in the positive shift direction is more detrimental than a systematic error in the negative direction.

\begin{acknowledgments}
The authors thank Dr.~Maciej Fidrysiak for insightful discussions. L.M., M.M.S. and G.G. acknowledge support by the projects PRIN2017 ``Quantum-2D” ID 2017Z8TS5B and PRIN2020 ``QT-FLUO`` ID 20207ZXT4Z of the Ministry for University and Research (MUR) of Italy. R.A. acknowledges support by the Swedish Research Council (VR) under the Project 2020-04945. This work was performed in part at Myfab Chalmers. The RIXS experimental data were collected at the beamline ID32 of the European Synchrotron (ESRF) in Grenoble (France), under the proposals HC-4620 (DOI: 10.15151/ESRF-ES-540330648) and IH-HC-3939 (DOI: 10.15151/ESRF-ES-1361861548).
\end{acknowledgments}

\bibliography{SGPGbiblio}
\end{document}